\documentclass{elsarticle}
\usepackage{graphics}
\usepackage{epsfig}
\usepackage{amsmath}
\usepackage{amsfonts}
\usepackage{amssymb}
\usepackage[latin1]{inputenc}
\usepackage[hmargin=2.5cm,vmargin=3cm]{geometry}
\usepackage[usenames]{color}

\title{Matching Pursuits with Random Sequential Subdictionaries}
\author{Manuel Moussallam$^1$,  Laurent Daudet$^{2,3}$ and Ga\"{e}l Richard$^1$  \\ \vspace{5mm} 
$^1$Institut Telecom - Telecom ParisTech - CNRS/LTCI  37/39, rue Dareau 75014 Paris, France \\ $^2$Institut Langevin - ESPCI ParisTech - Paris Diderot University - UMR7587 \\ 1, rue Jussieu 75238 PARIS CEDEX 05 \\
$^3$Institut Universitaire de France \\ contact: manuel.moussallam@telecom-paristech.fr}

\begin{document}
 \begin{abstract}
Matching pursuits are a class of greedy algorithms commonly used in signal processing, for solving the sparse approximation problem. They rely on an atom selection step that requires the calculation of numerous projections, which can be computationally costly for large dictionaries and burdens their competitiveness in coding applications. We propose using a non adaptive random sequence of subdictionaries in the decomposition process, thus parsing a large dictionary in a probabilistic fashion with no additional projection cost nor parameter estimation. A theoretical modeling based on order statistics is provided, along with experimental evidence showing that the novel algorithm can be efficiently used on sparse approximation problems. An application to audio signal compression with multiscale time-frequency dictionaries is presented, along with a discussion of the complexity and practical implementations.
\end{abstract}

\maketitle

\section*{keywords}
Matching Pursuits ; Random Dictionaries; Sparse Approximation; Audio Signal Compression

\section{Introduction}
\label{sec:Intro}
The ability to describe a complex process as a combination of few simpler ones is often critical in engineering. 
Signal processing is no exception to this paradigm, also known in this context as the sparse representation problem. 
Yet rather simple to expose, its combinatorial nature has led researchers to develop so many bypassing strategies that it can be considered a self-sufficient topic. 
Throughout these years of work, many additional benefits were found arising from the dimensionality reduction. Firstly, sparsity allows faster processing, and biologically inspired models \cite{Smith2005} suggest that the mammalian brain takes advantage of it. 
Secondly, it helps reducing both storage and broadcasting costs \cite{Ravelli2008}. Finally, it enables semantic characterization, since few significant objects carry most of the signals information.

The central problem in sparse approximation theory may be written as such: Given a signal $f$ in a Hilbert space $\mathcal{H}$ and a finite-size dictionary $\Phi = \{ \phi_{\gamma} \}$ of $M$ unit norm vectors ($\forall \gamma \in [1..M ]$ , $ \| \phi_{\gamma} \|^2 = 1$) in $\mathcal{H}$, called atoms, find the smallest expansion of $f$ in $\Phi$ up to a reconstruction error $\epsilon$ :
\begin{equation}
\min \| \alpha \|_0 \text{    s.t.    }   \| f - \sum_{\gamma = 1}^{M} \alpha_{\gamma} \phi_{\gamma} \|^2 \leq \epsilon
\label{eq:problem_combi}
\end{equation}
where $\| \alpha \|_0$ is the number of non-zero elements in the sequence of weights $\{ \alpha_{\gamma} \}$. Equivalently, one seeks the subset of indexes $\Gamma^{m} = \{ \gamma_n \}_{n = 1.. m}$ of atoms of $\Phi$ and the corresponding $m$ non zero weights $\{ \alpha_{\gamma_n} \}$ solving:
\begin{equation}
\min m \text{    s.t.    }  \| f - f_m \|^2 \leq \epsilon
\end{equation}
where $f_m = \sum_{n = 1}^{m} \alpha_{\gamma_n} \phi_{\gamma_n}$ is called a $m$-term approximant of $f$.

A corollary problem is also defined, the sparse recovery problem, where one assumes that $f$ has a sparse support $\Gamma^{m}$ in $\Phi$ ($m$ being much smaller than the space dimension) and tries to recover it (from noisy or fewer measurement than the signals dimension). In the sparse approximation problem, no such assumption is made and one just seeks to retrieve the best $m$-term approximant of $f$ in the sense of minimizing the reconstruction error (or any adapted divergence measure). Many practical problems benefit from this modeling, from denoising \cite{Fevotte2008,Chen98atomicdecomposition} to feature extraction and signal compression \cite{FiguerasiVentura2006,Ravelli2008}.

Unfortunately, problem (\ref{eq:problem_combi}) is NP-hard and an alternate strategy needs to be adopted. A first class of existing methods are based on a relaxation of the sparsity constraint using the $l_1$-norm instead of the non-convex $l_0$ \cite{Tibshirani94LASSO,Chen98atomicdecomposition}, thus solving a quadratic program. Alternatively, greedy algorithms can produce (potentially suboptimal) solutions to problem (\ref{eq:problem_combi}) by iteratively selecting atoms in $\Phi$.

Matching Pursuit (MP) \cite{Mallat_TSP1993} and its variants \cite{Pati93orthogonalmatching,Jaggi1998,Daudet2006,Christensen2007,Blumensath_GP2008} are instances of such greedy algorithms. A comparative study of many existing algorithms can be found in \cite{Dymarski2011}. 
MP-like algorithms have simple underlying principles allowing intuitive understanding, and efficient implementations can be found \cite{Krstulovic_ICASSP2006,Mailhe2009_LocOMP}. Such algorithms
construct an approximant $f_n$ after $n$ iterations by alternating two steps:
\begin{description}
 \item[Step 1] Select an atom $\phi_{\gamma_n} \in \Phi$ and append it to the support $\Gamma^n = \Gamma^{n-1} \bigcup \gamma_n$
 \item[Step 2] Update the approximation $f_{n}$ accordingly.
\end{description}
until a stopping criterion is met or a perfect decomposition is found (i.e $f_n = f$). This two-step generic description defines the class of General MP algorithms \cite{Gribonval2006}.


When designing the dictionary, it is important that atoms locally ressemble the signal that is to be represented. This correlation between the signal and the dictionary atoms ensures that greedy algorithms select atoms that remove a lot of energy from the residual).
For some classes of signals (e.g. audio), the variety of encountered waveforms implies that a large number of atoms must be considered.
Eventually, one is interested in finding sparse expansions of signals in large dictionaries at reasonable complexity. 

Methods addressing sparse recovery problems have benefited from random dictionary design properties (e.g with MP-like algorithms \cite{Tropp_OMP2007, Blumensath_GP2008}). Randomness in this context is used for spreading information that is localized (on a few non-zero coefficients in a large dictionary) among fewer measurement vectors.

In the sparse approximation context though, the use for randomness is less obvious. Much effort has been done on designing structured dictionaries that exhibit good convergence pattern at reasonable costs \cite{Ravelli2008,Daudet2006,Jost04tree-basedpursuit}.
Typical dictionaries with limited size are based on Time-Frequency transforms such as window Fourier transforms (also referred to as Gabor dictionaries) \cite{Mallat_TSP1993,Gribonval1996},  Discrete Cosine \cite{Ravelli2008}, wavelets \cite{Smith2005} or unions of both \cite{Daudet2006}. Such dictionaries are made of atoms that are localized in the time-frequency plance, which gives them enough flexibility to represent complex non stationary signals. However the set of considered localization reflects, in practice, a compromise.

Considering all possible localizations is, in theory, possible for discrete signals and dictionaries, however it yields very large and redundant dictionaries which raises computational issues. The standard dictionary design \cite{Mallat_TSP1993} considers a fixed subset of localizations is arguably a good option but it can never be optimal for all possible signals. 
Adapting the subset to the signal is then possible, but it introduces additional complexity. 
Finally, one may consider using multiple random subsets, perform multiple decompositions and average them in a post-processing step \cite{Ferrando2000}.

In this paper, an other option is considered: the use of randomly varying subdictionaries at each iteration during a single decomposition. The idea is to avoid the additional complexity of adapting the dictionary to the signal, or of having to perform multiple decompositions, while still improving on the fixed dictionary strategy. 

A parallel can be traced with quantization theory. Somehow, limiting dictionary size amounts to discretizing its atoms parameter space. This quantization introduces noise in the decomposition (i.e error due to dictionary-related model mismatch). Adaptive quantization reduces the noise and multiple quantization allow to average the noise out. But the proposed technique is closer to the dithering technique \cite{Zamir92} and more generally to the stochastic resonance theory, which shows how a moderate amount of added noise can increase the behavior of many non linear systems \cite{Gammaitoni1998}.
 
The proposed method only relies on a slight modification of step 1 and can therefore be applied to many pursuit algorithms. In this work, we focus on the approximation problem. Experiments on real audio data demonstrate the potential benefits in signal compression and particularly in low-bit-rate audio coding. It should be emphasized that, as such, it is not well suited for sparse recovery problems such as in the compressed sensing scheme \cite{Donoho2006}. Indeed, searching in random subsets of $\Phi$ may burden the algorithm ability to retrieve true \textit{true} sparse support of a signal in $\Phi$.

The rest of this paper is organized as follows. Section \ref{sec:greedyAlgos} recalls the standard Matching Pursuit algorithm and some of its variants. Then, we introduce the proposed modification of Step 1 in Section \ref{sec:theoryRMP} and expose some theoretical justifications based on a probabilistic modeling of the decomposition process. A practical application is presented in Section \ref{sec:RMP_Sounds} as a proof of concept with audio signals. Extended discussion on various aspects of the new method is provided in Section \ref{sec:discussion}.

\section{Matching Pursuit for Sparse Signal Representations}
\label{sec:greedyAlgos}

\subsection{Matching Pursuit Framework}
\begin{figure*}
 \centering
 \includegraphics[width=16cm]{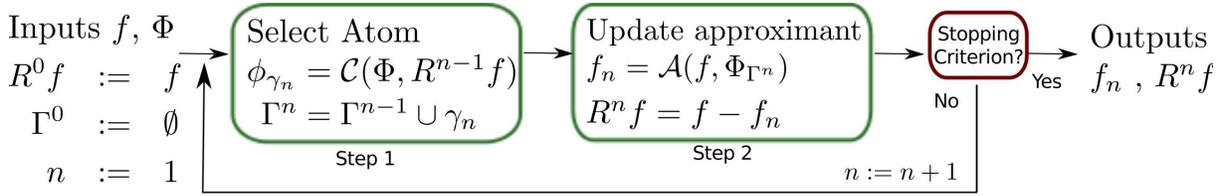}
 \caption{\textit{Block diagram for General MP algorithms. An input signal $f$ is decomposed onto a dictionary $\Phi$ by iteratively selecting atoms $\phi_{\gamma_{n}}$ (Step 1) and updating an approximant $f_n$ (Step 2).}}
 \label{fig:Greedy_schema}
\end{figure*}

The Matching Pursuit (MP) algorithm \cite{Mallat_TSP1993} and its variants in the General MP family \cite{Gribonval2006} all share the same underlying iterative two-step structure, as described by Figure \ref{fig:Greedy_schema}.
Specific algorithms differ in the way they perform the atom selection criteria $\mathcal{C}$ (Step 1), and the approximant construction $\mathcal{A}$ (Step 2). Standard MP as originally defined in \cite{Mallat_TSP1993} is given by:
\begin{eqnarray}
\label{eq:function_C}
 \mathcal{C}(\Phi  , R^{n} f)  &=& \arg \max_{ \phi_{\gamma} \in \Phi} |\langle R^{n}f , \phi_{\gamma} \rangle | \\
 \mathcal{A}(f , \Phi_{\Gamma^{n}} ) &=& \sum_{i=0}^{n-1} \langle R^{i}f , \phi_{\gamma_i} \rangle \phi_{\gamma_i}  
\end{eqnarray}
where $R^nf$ denotes the residual at iteration $n$ (i.e $R^n f = f - f_n$).
Orthogonal Matching Pursuit (OMP) \cite{Pati93orthogonalmatching} is based on the same selection criteria, but the approximant update step ensures orthogonality between the residual and the subspace spanned by the already selected atoms $\mathcal{V}_{\Gamma^n} = \text{span}\{ \phi_{\gamma_n}\}_{\gamma_n \in \Gamma^n}$
\begin{eqnarray}
 \mathcal{A}_{OMP}(f , \Phi_{\Gamma^{n}} ) &=&  P_{\mathcal{V}_{\Gamma^n}} f
\end{eqnarray}
where $P_{\mathcal{V}}$ is the orthonormal projector onto $\mathcal{V}$.

The algorithm stops when a predefined criterion is met, either an approximation threshold in the form of a \textit{Signal-to-Residual Ratio} (SRR):
\begin{equation}
 SRR(n) = 10 \log \frac{\|f_n\|^2}{\| f - f_n\|^2}
\end{equation}
or a bit budget (equivalently a number of iterations) in coding applications.

\subsection{Weak MP}
Often, the size of $\Phi$ prevents the use of the criterion $\mathcal{C}$ as defined in \ref{eq:function_C}, especially for infinite-size dictionaries. Instead, one can satisfy a weak selection rule $\mathcal{C}_{weak}(\Phi  , R^{n} f) = \phi_{\gamma^{weak}}$ such that:
\begin{equation}
\label{eq:criteria_weak}
|\langle R^{n}f , \phi_{\gamma^{weak}}\rangle | \geq t_{n} \sup_{ \phi_{\gamma} \in \mathbf{\Phi}} |\langle R^{n}f , \phi_{\gamma} \rangle |
\end{equation}
with $0 < t_{n} \leq 1$. In practice, a weak selection step is implemented by limiting the number of atomic projections in which the maximum is searched. This is equivalently seen as a subsampling of the large dictionary $\Phi$.
Temlyakov \cite{Temlyakov2002} proved the convergence when $\sum \frac{t_n}{n} < \infty$ and stability properties have been studied by Gribonval \textit{et al} in \cite{Gribonval2006}. 

Let us denote $\Phi = \{ \phi_{\gamma} \}$ a large multiscale time-frequency dictionary with atom parameter $\gamma$ living in $S \times U \times \Xi$ a finite subset of $\mathbb{R}^{+} \times \mathbb{R}^2$. $S$ denotes the set of analysis scales, $U$ is the set of time indexes and $\Xi$ the set of frequency indexes (see for example \cite{Ravelli2008}). 
Computing $\mathcal{C}(\Phi  , R^{n-1} f)$ requires, in the general case, the computation of all atom projections $\langle R^{n-1}f , \phi_{\gamma} \rangle$ which can be prohibitive. 
A weak strategy consists in computing only a subset of these projections, thus using a coarse subdictionary $\Phi_{\mathcal{I}} \subset \Phi$, which can be seen as equivalent to a subsampling of the parameter space $S \times U \times \Xi$. 
For instance, time indexes can be limited to fractions of the analysis window size.

\begin{figure}
 \centering
 \includegraphics[width=8cm]{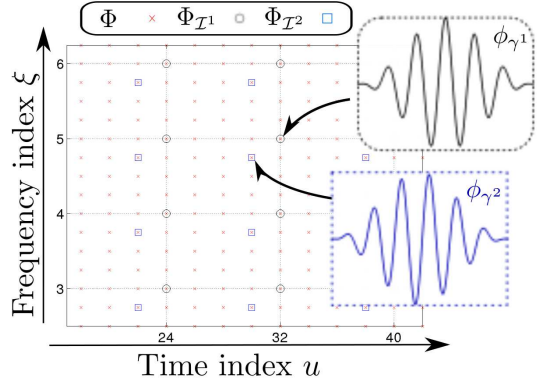}
 \caption{\textit{Time Frequency grids defined by $\Phi$ a redundant monoscale time-frequency dictionary, $\Phi_{\mathcal{I}^1}$ and $\Phi_{\mathcal{I}^2}$, two subsets of $\Phi$. Atoms from $\Phi_{\mathcal{I}^2}$ are almost the same as $\Phi_{\mathcal{I}^1}$ but with an additional time and/or frequency offset.}}
 \label{fig:grids2}
\end{figure}

Reducing the size of the dictionary is interesting for coding purposes. Actually, the smaller the atom parameter space, the cheaper the encoding of atom indexes. Figure \ref{fig:grids2} shows an example of such dictionary subsampling in a time-frequency plane. Each point represents the centroid of a time-frequency atom $\phi_{\gamma}$, black circles corresponds to a coarse subsampling of the parameter space and constitute the subdictionary $\Phi_{\mathcal{I}^1} \subset \Phi$. The full dictionary $\Phi$ figured by the small red crosses, is 16 times bigger than $\Phi_{\mathcal{I}^1}$.

However, the choice of subsampling has consequences on the decomposition. 
Figure \ref{fig:grids2} pictures an example of a different coarse subdictionary defined by another index subset $\Phi_{\mathcal{I}^2} \subset \Phi$. For most signals, the decomposition will be different when using $\Phi_{\mathcal{I}^1}$ or $\Phi_{\mathcal{I}^2}$. Moreover, there is no easy way to guess which subdictionary would provide the fastest and/or most significant decomposition. Using one of these subdictionaries instead of $\Phi$ implies that available atoms may be less suited to locally fit the signals. MP would thus create energy in the time frequency plane where there is none in the signal, an artifact known as dark energy \cite{Sturm2008}.

\subsection{Higher Resolution and Locally-Adaptive Matching Pursuits}
To tackle this issue, high resolution methods have been studied. First, a modification of the atom selection criterion has been proposed by Jaggi \textit{et al} \cite{Jaggi1998} that not only takes energy into account but also local fit to the signal. Other artifact preventing methods have also been introduced (e.g for pre-echo \cite{Ravelli2008} or dark energy \cite{Sturm2010}). 
A second class of bypassing strategies are locally-adaptive pursuits. An atom is first selected in the coarse subdictionary $\Phi_{\mathcal{I}}$, then a local optimization is performed to find a neighboring maximum in $\Phi$. Mallat \textit{et al} proposed a Newton method \cite{Mallat_TSP1993}, Goodwin \textit{et al} focused on phase tuning \cite{Goodwin1999}, Gribonval tuned a chirp parameter \cite{Gribonval_Chiprs2001} and Christensen \textit{et al} addressed both phase and frequency \cite{Christensen2007,Sturm2010a}. An equivalent strategy is defined for gammachirps in \cite{Lewicki98codingtime-varying}. 
Locally-adaptive methods in time-frequency dictionaries implement a form of time and / or frequency reassignment of an atom selected in a coarse subdictionary. 

These strategies yield representations in the large dictionary at a reduced cost. In many applications, the faster residuals energy decay compensates the slight computational overhead of the local optimization. In a previous work \cite{Moussallam2011}, we emphasized the usefulness of locally-adaptive methods for shift-invariant representation and similarity detection in the transform domain.
Though, the resulting atoms are from the large dictionary, i.e their indexes are more costly to encode. Moreover, these method require an additional parameter estimation step that may increase the overall complexity.


\subsection{Statistics and MP}
Different approaches have been proposed to enhance MP algorithms using statistics. 
Ferrando \textit{et al} \cite{Ferrando2000} were (to our knowledge) the first to propose to run multiple sub-optimal pursuits and retrieve meaningful atoms in an \textit{a posteriori} averaging step. A somehow similar approach is introduced in \cite{Durka2001_StochasticMP} where the authors emphasize a lack of precision of their representation due to the structure of the applied dictionary. To avoid these decomposition artifacts, they follow a Monte-Carlo like method, where the set of atom parameter is randomly chosen before each decomposition followed by an averaging step. 

Elad \textit{et al} \cite{Elad2009} used the same paradigm for support recovery, taking advantage of many suboptimal representations instead of a single one of higher precision. 
Similar approaches are adopted within a Bayesian framework in \cite{Schniter2008,Zayyani2009}, although the subset selection is a byproduct of the choice of the prior. Their work is related to the branch of compressed sensing that uses pursuits and random measurement matrices to retrieve sparse supports of various classes of high dimensional signals \cite{Tropp_OMP2007}. A recent work by Divekar \textit{et al} \cite{Divekar2010} proposed a form of probabilistic pursuit.
Such work differs from ours in the sense that it is signal-adaptive and tuned for the support recovery problem.



\section{Matching Pursuits with Sequences of Subdictionaries}
\label{sec:theoryRMP}
\subsection{Proposed new algorithm}
\label{subsec:newAlgo}

This paper presents a modification of MP which consists in changing the subdictionary at each iteration. 
Let us define a sequence $\mathbf{I} = \{ \mathcal{I}^n \}_{n \in [0..m-1]}$ of length $m$. Each $\mathcal{I}^n$ is a set of indexes of atoms from $\Phi$. 
At iteration $n$, the algorithm can only select an atom in the subdictionary $\Phi_{\mathcal{I}^n} \subset \Phi$. We call such algorithms Sequential Subdictionaries Matching Pursuits (SSMP).

\begin{figure*}
 \centering
 \includegraphics[width=16cm]{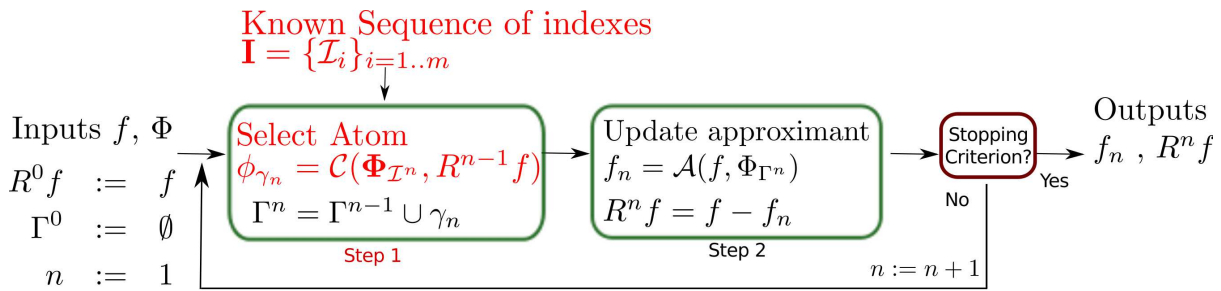}
 \caption{\textit{Block diagram of Sequential Subdictionaries Pursuits (SSMP). Step 1 is modified so as to have the search take place in a different subdictionary at each iteration. The dictionary subsampling is controlled by a fixed pre-defined sequence $\mathbf{I}$.}}
 \label{fig:RandomSchema}
\end{figure*}

This method is illustrated by Figure \ref{fig:RandomSchema}. When the sequence $\mathbf{I}$ is random, we call such algorithms Random Sequential Subdictionaries (RSS) Pursuits. The proposed modification can be applied to any algorithm from the General MP family (RSS MP, RSS OMP, etc..).
After $m$ iterations, the algorithm outputs an approximant of the form (\ref{eq:approximantRSSMP}) 
\begin{equation} 
\label{eq:approximantRSSMP}
 f_m = \sum_{n=0}^{m-1} \alpha_n \phi_{\gamma_n}^{\mathcal{I}^n}
\end{equation}
where $\phi_{\gamma_n}^{\mathcal{I}^n} \in \Phi_{\mathcal{I}^n}$.

SSMP can be seen as a Weak General MP algorithm as described in \cite{Gribonval2006}. Indeed, it amounts to reducing the atom selection choice to a subset of $\Phi$, which defines the \textit{weak} selection rule (\ref{eq:criteria_weak}). The originality of this work is the fact that we change the subdictionary at each iteration while it is usually fixed during the whole decomposition. The sequence of subdictionaries is known in advance and is thus signal independent. 

The sequence $\mathbf{I}$ might be built such that $\lim_{m \rightarrow \infty}\bigcup_{n=0}^{m-1} \Phi_{\mathcal{I}^n} = \Phi$, although in this case, the modified pursuit is not equivalent to a pursuit using the large dictionary $\Phi$. 
However, it allows the selection of atoms from $\Phi$ that were not in the initial subdictionary $\Phi_{\mathcal{I}^0}$. 

\subsection{A first example}
\label{sec:first_example}
Let $\Phi$ be a Gabor dictionary. Each atom in $\Phi$ is a Gabor atom that can be parameterized by a triplet $(s,u,\xi) \in S \times U \times \Xi$ defining its scale, time and frequency localization respectively. The continuous-time version of such atom is given by:
\begin{equation}
 \phi_{ s , u , \xi}(t) = \frac{1}{\sqrt{s}} g \left(\frac{t - u}{s} \right) e^{i\xi (t-u)}
\end{equation}
where $g$ can be a Gaussian or Hann window. A discrete Gabor dictionary is obtained by sampling the parameter space $S \times U \times \Xi$ (and the window $g$).
Let $N$ be the signal dimension and $K$ be the number of Gabor scales ($s_1, s_2,..,s_K$). For each scale $s_k$, $u$ takes integer values between $0$ and $N$ and $\xi$ from $0$ to $s_k/2$. The total size of $\Phi$ in this setup is $M = \sum_{k=1}^K \frac{s_k N}{2}$ atoms.

Let $f \in \mathbb{R}^N$ be a $m$-sparse signal in $\Phi$, meaning that $f$ is the sum of $m$ (randomly chosen) components from the full dictionary $\Phi$. 
We are interested in minimizing the reconstruction error $\epsilon = \frac{\|f - f_{n} \|^2}{\|f\|^2}$ given the size constraint $n = m/2$.

Let $\Phi_{\mathcal{I}^0}$ be a subdictionary of $\Phi$ defined by selecting in each scale $s_k$ only atoms at certain time localizations. Subsampling the time axis by half the window length is standard in audio processing (i.e. $\forall u$  $\exists p \in [0 , \lfloor2N/s_k\rfloor -1], \text{  s.t. } u = p.s_k/2$). $\Phi_{\mathcal{I}^0}$ can be seen as a subsampling of $\Phi$. The size of $\Phi_{\mathcal{I}^0}$ in this setup is down to $K \times N$ atoms. MP using $\Phi_{\mathcal{I}^0}$ during the whole decomposition is a special case of SSMP where all the subdictionaries are the same. We label this algorithm Coarse MP.

Now let us define a pseudo-random sequence of subdictionaries $\Phi_{\mathcal{I}^n}$ that are translated versions of $\Phi_{\mathcal{I}^0}$. Each $\Phi_{\mathcal{I}^n}$ has the same size than $\Phi_{\mathcal{I}^0}$ but its atoms are parameterized as such: $\forall u$  $\exists p \in [0 , \lfloor2N/s_k\rfloor -1], \text{  s.t. } u = p.s_k/2 + \tau_k^n$ where $\tau_k^n \in [-s_k/4 , s_k/4]$ is a translation parameter different for each scale $s_k$ of each subdictionary $\Phi_{\mathcal{I}^n}$.  Each $\Phi_{\mathcal{I}^n}$ can also be seen as a different subsampling of $\Phi$. The size of each subdictionary $\Phi_{\mathcal{I}^n}$ is thus also $K \times N$. MP using the  sequence of subdictionaries $\Phi_{\mathcal{I}^n}$ is labeled Random SSMP (RSS MP).

\begin{figure}
 \centering
 \includegraphics[width=8cm]{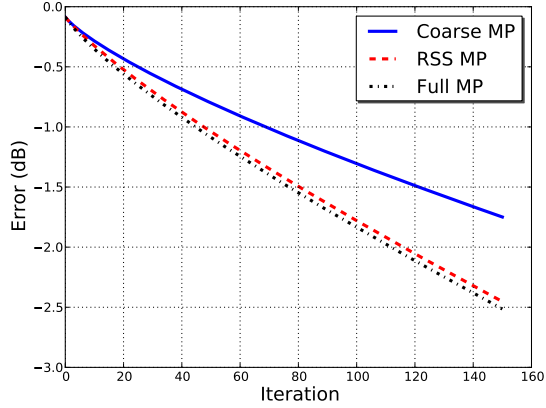}
 \caption{Evolution of normalized approximation error $\epsilon$ with Matching Pursuit using a fixed subdictionary (Coarse MP), a pseudo random sequence of subdictionaries (RSS MP) and the full original dictionary (Full MP). Results averaged over 1000 runs.}
 \label{fig:ReconstructionError}
\end{figure}

Finally, a MP using the full dictionary $\Phi$ during the whole process is labeled Full MP. We compare the reconstruction error achieved with these three algorithms (i.e Coarse MP, RSS MP and Full MP).

Figure \ref{fig:ReconstructionError} gives corresponding results with the following setting: $N = 10000$, $m = 300$, $S = [32,128,512]$ and thus $M = 840000$, averaged over 1000 runs. Although RSS MP is constrained at each iteration to choose only within a limited subset of atoms, it exhibit an error decay close to the one of Full MP.

If $f$ was exactly sparse in $\Phi_{\mathcal{I}^0}$, then keeping the fixed subdictionary (Coarse MP) would provide a faster decay than using the random sequence (RSS MP). However in our setup, such case is unlikely to happen. As stressed by many authors cited above, the choice of a fixed subdictionary is often suboptimal with respect to a whole class of signals, as it may not have the basic invariants one may expect from a good representation (for instance, the shift invariance). A matter of concern is now to determine when RSS MP performs better than Coarse MP in terms of approximation error decay.

In the next subsection we make use of order statistics to model the behavior of the two strategies (i.e. MP with fixed coarse subdictionary and MP with varying subdictionaries) depending on the initial distribution of projections. 

\subsection{Order Statistics}
\label{subsec:order_stats}
We need to introduce a few tools from the order statistics theory. The interested reader can refer for example to \cite{Nagaraja1990,Gupta1972} for more details.
Let $z_1, z_2, .. z_n$ be $n$ i.i.d samples drawn from a continuous random variable $Z$ with probability density $f_Z$ and a distribution function $F_Z$. Let us denote $Z_{1:n} , Z_{2:n},.. , Z_{n:n}$ the order statistics. The random variable $Z_{i:n}$ represents the $i^{th}$ smallest element of the $n$ samples. The probability density of $Z_{i:n}$ is denoted $f_{i:n}^Z$ and is given by:
\begin{equation}
 \label{eq:i_orderStat_pdf}
f_{i:n}^Z(z) = \frac{n!}{(n-i)! (i-1)! } F_Z(z)^{i-1}  f_Z(z)  (1- F_Z(z))^{n-i}
\end{equation}
The density of extremum values is easily derived from this equation, in particular the maximum value of the sequence has density:
\begin{equation}
 f_{n:n}^Z(z) = n F_Z(z)^{n-1}  f_Z(z) 
\end{equation}
The moments of the order statistics are given by the formula:
\begin{equation}
\label{eq:momentFormula}
 \mu_{i:n}^{(m)} = \mathbb{E}(Z_{i:n}^m) = \int_{-\infty}^{\infty} z^m f_{i:n}^Z(z) dz
\end{equation}
And the variance is denoted $\sigma^2_{i:n}$. For convenience we will write the expectation $\mu_{i:n} = \mu_{i:n}^{(1)}$. 

\subsubsection{Order statistics in an MP framework}
Given the greedy nature of MP, a statistical modeling of the series of projection maxima gives meaningful insights about the overall convergence properties. Let $f$ be a signal in $\mathbb{R}^N$. At the first iteration, the projection of the residual $R^{0}f = f$  over a complete dictionary $\Phi$ of $M$ unit normed atoms $\{ \phi_i\}_{i \in [1,M]}$ ($M > N$) is given by:
\begin{equation}
  \forall i \in [1..M], \alpha_i = \langle R^{0} f , \phi_i \rangle
\end{equation}
Let us denote $z_i = |\alpha_i|$ and consider them as $M$ i.i.d samples drawn from a random variable $Z$ living in $[0, \sqrt{\|f\|^2}]$. Note that such an assumption holds only if one considers that atoms in $\Phi$ are almost pairwise orthogonals, i.e that $\Phi$ is quasi-incoherent. MP selects the atom $\phi_{\gamma_0} = \arg \max z_i$ whose weights absolute value is given by the $M^{th}$ order statistics of $Z$: $Z_{M:M}$.
\begin{equation}
 \label{eq:maxMP_KorderStat}
| \alpha_{\gamma_0} | = \max_{\phi_{\gamma} \in \Phi} | \langle R^{0}f , \phi_{\gamma} \rangle | = Z_{M:M} 
\end{equation}
Standard MP constructs a residual by removing this atoms contribution and iterating. Hence, weights of selected atoms remains independent and at iteration $n$ the $M-n$-th order statistics of $Z$ describes the selected weight: 
\begin{equation}
 \label{eq:maxMP_KorderStat_itn}
| \alpha_{\gamma_n} | = \max_{\phi_{\gamma}\in \Phi} | \langle R^{n}f , \phi_{\gamma} \rangle | = Z_{M-n:M} 
\end{equation}
the energy conservation in MP allows to derive (\ref{eq:MP_energyConservation}).
\begin{equation}
\label{eq:MP_energyConservation}
 \| f\|^2 = \|R^nf \|^2 + \sum_{i=0}^{n-1} |\alpha_{\gamma_i}|^2
\end{equation}
Combining (\ref{eq:MP_energyConservation}) and (\ref{eq:maxMP_KorderStat_itn}) and taking the expectation, one gets (\ref{eq:residual_MP_itn})
\begin{equation}
\label{eq:residual_MP_itn}
 \mathbb{E}(\| R^n f \|^2) =  \| f \|^2 -  \sum_{i=0}^{n-1} \mu_{M-i:M}^{(2)}
\end{equation}
We recognize the second order moment of $\mu_{M-i:M}^{(2)} = \sigma_{M-i:M}^{2} + \mu_{M-i:M}^{2}$. Similarly, from (\ref{eq:MP_energyConservation}) we can derive the variance of the estimator:
\begin{equation}
\label{eq:residual_MP_itn_variance}
 Var(\| R^n f \|^2) =  \sum_{i=0}^{n-1} \sum_{j=0}^{n-1} cov ( Z_{M-i:M}^{2} , Z_{M-j:M}^{2})
\end{equation}
Given a distribution model for $Z$, Equations (\ref{eq:residual_MP_itn}) and (\ref{eq:residual_MP_itn_variance}) provide mean and variance estimates of the residuals energy decay with a standard MP on a quasi-incoherent dictionary $\Phi$.

\subsubsection{Redrawing projection coefficients: changing the dictionaries}
\label{subsec:redrawing}
The idea at the core of the new algorithm described in section \ref{subsec:newAlgo} is to draw a new set of projections at each iteration by changing the dictionary in a controlled manner. Let us now assume that we know a sequence of complete quasi-incoherent dictionaries $\{\Phi_i\}_{i \in [0,n]}$. 
Let us make the additional assumptions that the projection of $f$ has the same distribution in all these dictionaries.

At the first iteration, the process is similar, so the atom $\phi_{\gamma_0}$ in $\Phi_0$ is selected as $\arg \max_{\phi_{\gamma} \in \Phi_0} |\left\langle R^0f , \phi_{\gamma} \right\rangle|$ with the same weight as (\ref{eq:maxMP_KorderStat}). After subtracting the atom we have a new residual $R^1f$. The trick is now to search for the most correlated atom in $\Phi_1$. The absolute values of the corresponding inner products define a new set of $M$ i.i.d samples $z_i^1$:
\begin{equation}
  \forall i \in [1..M], z_i^1 = | \alpha_i | = |\langle R^{1} f , \phi_i \rangle|
\end{equation}
Let us assume that the $z_i^1$ samples have the same distribution law than the $z_i$, except that the sample space is now smaller because $R^1f$ has less energy than $f$. This means the new random variable $Z1$ is distributed like $Z \times \frac{\| R^1f \|}{\| f \|}$. Let us denote $Z1_{M:M}$ the $M$-th order statistic for $Z1$, its second moment is:
\begin{equation}
 \mathbb{E}(Z1_{M:M}^2) = \mathbb{E}(Z_{M:M}^2) . \mathbb{E}( \frac{\| R^1f \|^2}{\| f \|^2}) + cov ( Z_{M:M}^2 , \frac{\| R^1f \|^2}{\| f \|^2})
\end{equation}
and we know that $\| R^1f \|^2 = \|f\|^2 - Z_{M:M}^2 $ hence:
\begin{equation}
 cov ( Z_{M:M}^2 , \| R^1f \|^2) = - cov ( Z_{M:M}^2 , Z_{M:M}^2) = (\mu_{M:M}^{(2)})^2 - \mu_{M:M}^{(4)}
\end{equation}
which leads to:
\begin{eqnarray*}
 \mathbb{E}(Z1_{M:M}^2) &=& \mu_{M:M}^{(2)} .(1 - \frac{\mu_{M:M}^{(2)}}{\| f \|^2}) + \frac{1}{\| f \|^2}\left((\mu_{M:M}^{(2)})^2 - \mu_{M:M}^{(4)}\right) \\
 &=& \mu_{M:M}^{(2)} - \frac{\mu_{M:M}^{(4)}}{\| f \|^2}
\end{eqnarray*}
Since the new residual $R^2f$ has energy $\| R^2f \|^2 = \| R^1f \|^2 - Z1_{M:M}^2 $, taking the expectation:
\begin{equation}
\label{eq:residual_RSSMP_it2}
 \mathbb{E}(\|R^2f\|^2) =\|f\|^2 - 2\mu_{M:M}^{(2)} + \frac{\mu_{M:M}^{(4)}}{\| f \|^2} 
\end{equation}

and we see higher moments of the $M$-th order statistic appear in the residuals energy decay formula. At the $n$-th iteration of this modified pursuit, we end up with a rather simple formula (\ref{eq:residual_RSSMP_itn}) where higher moments of the $M$-th order statistic of $Z$ appear.
 \begin{equation}
\label{eq:residual_RSSMP_itn}
 \mathbb{E}(\| R^n f \|^2) = \| f \|^2 + \sum_{i=1}^{n} (-1)^{j} {n \choose i}  \frac{\mu_{M:M}^{(2i)}}{ \| f \|^{2(i-1)}}
\end{equation}
Similarly, a variance estimator can be derived that only depends on the moments of the highest order statistic:
 \begin{equation}
\label{eq:residual_RSSMP_itn_variance}
 Var(\| R^n f \|^2) = \sum_{i=1}^{n} \sum_{j=1}^{n} (-1)^{j+i} {n \choose i} {n \choose j} \frac{cov(Z_{M:M}^{(2i)}, Z_{M:M}^{(2j)})}{ \| f \|^{2(i+j-1)}}
\end{equation}

\subsection{Comparison with simple models}
\label{sec:simple_models}
Let us now compare the behavior of the new redrawn samples strategy to the one where they are fixed. 

The relative error $\epsilon(n) = 10 \log \frac{\|R^nf\|^2}{\|f\|^2}$ after $n$ iterations is computed for both strategies. Knowing the probability density function (pdf) $f_Z$, Equations (\ref{eq:residual_MP_itn}) to (\ref{eq:residual_RSSMP_itn_variance}) provide closed-form mean and variance estimate of $\epsilon(n)$ for both strategies. For example if $Z \sim \mathcal{U}(0,1)$,  its order statistics follow a beta distribution: $Z_{k:M} \sim Beta(k , M+1-k)$ and all moments $\mu_{M-k:M}^{(m)}$ can be easily found. 

\begin{figure*}
\centering
  \includegraphics[width=16cm]{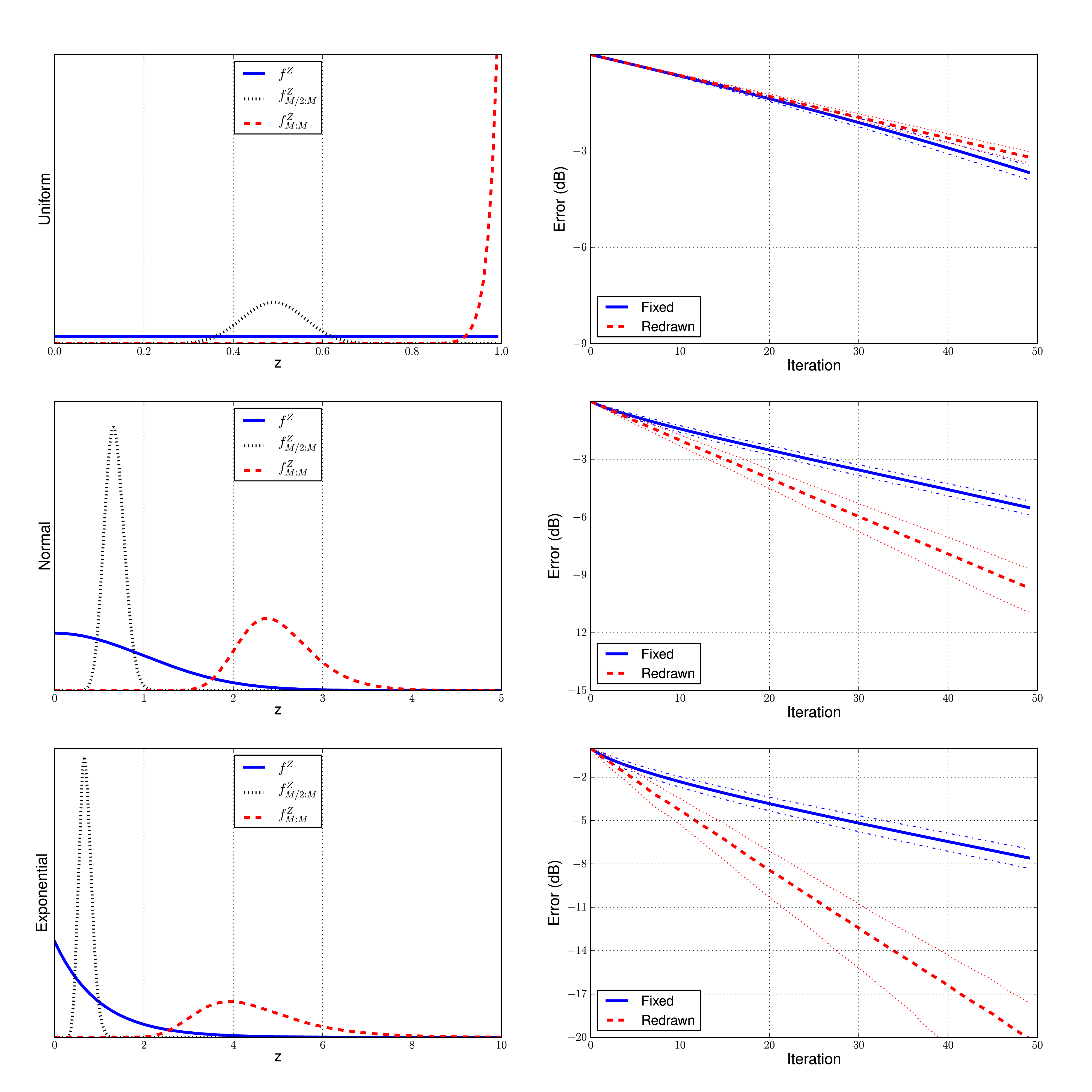}

\caption{Left : Pdf of $Z$ and pdfs for the maximum and the $M/2$-th order statistic of $Z$ for uniform, normal ($\sigma = 1$) and exponential ($\mu = 1$) distributions. Right: relative error decay (mean and variances) as a function of iteration number. $M=100$}
\label{fig:DistribResults}
\end{figure*}

A graphical illustration of potential of the new strategy is given by Figure \ref{fig:DistribResults}. For three different distributions of $Z$, namely uniform, normal (actually half normal since no negative values are considered) and exponential. The density function of two order statistics of interest (i.e the maximum $f_{M:M}^Z$, and the $M/2$-th element $f_{M/2:M}^Z$) is plotted. These two elements combined provide an insight on how fast the expected value of selected weight will decline using the fixed strategy. 

Some comments can be made:
\begin{itemize}
 \item In the uniform case: one might still expect to select relatively large coefficients (i.e \textit{good} atoms) after $M/2$ iterations with the standard strategy. In this case, redrawing the samples (i.e changing the subdictionary) appears to be detrimental to the error decay rate. 
 \item For normally and exponentially distributed samples, the expected value of a coefficient selected in a fixed sequence drops more quickly towards small values (i.e there are relatively fewer large values). Relative error profiles indicate that selecting the maximum of redrawn samples can prove beneficial in terms of minimizing the reconstruction error. It is promising for sparse approximation problems such as lossy compression. 
\end{itemize}

In practice, the uniform model is not well suited for sparse approximation problems. It basically indicates that the chosen dictionary is poorly correlated with the signal (e.g. white noise in Dirac dictionaries). Exponential and normal distributions, on the opposite, are closer to practical situations where the signal is assumed sparse in the dictionary, with additional measurement noise such as the one presented in section \ref{sec:first_example}. In these cases, changing the subdictionaries seems to be beneficial. 

Let us again stress that we have only presented a model for error decays with two strategies and a set of strong assumptions. It is not a formal proof of a guaranteed faster convergence. However, it gives insight on when the proposed new algorithm may be useful and when it may not, along with estimators of the decay rates when a statistical modeling of signal projections on subdictionaries is available. To our knowledge, such modeling was not proposed before for MP decompositions.

\section{Matching Pursuit with Random Time Frequency Subdictionaries for audio compression}
\label{sec:RMP_Sounds}

\subsection{Audio compression with union of MDCT dictionaries}
Sparse decompositions using Matching Pursuit have been proven by Ravelli \textit{et al} \cite{Ravelli2008} to be competitive for low bit-rate audio compression. In spite of Gabor dictionaries, they use the Modified Discrete Cosine Transform (MDCT) \cite{Princen1987} that is at the core of most transform coders (e.g. MPEG1-Layer III \cite{MP3-ISO92}, MPEG 2/4 AAC). Let us recall the main advantage of RSSMP compared to Coarse MP:  reconstruction error can be made smaller with the same number of iterations. If the random sequence of subdictionaries is already known by both coder and decoder then the cost of encoding each atom is the same in both cases.
\begin{figure*}
 \centering
 \includegraphics[width=16cm]{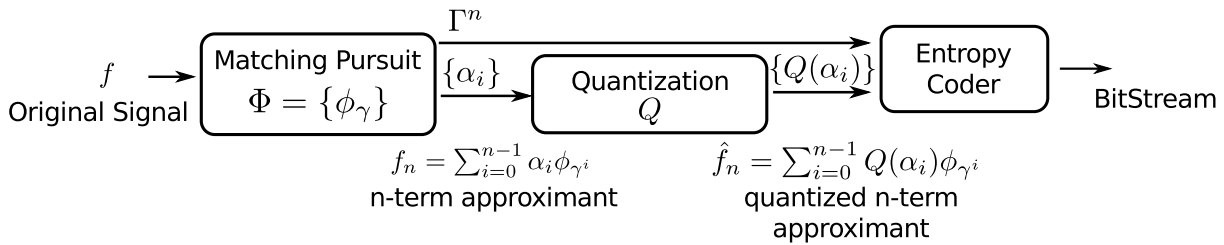}
 \caption{Block diagram of encoding scheme: each atom selected by the pursuit algorithm is transmitted after quantizing its associated weight. A simple entropy coder then yields the bitstream}
 \label{fig:Encoding}
\end{figure*}

For this proof of concept, a simplistic coding scheme is used as presented in Figure \ref{fig:Encoding}. For each atom in the approximant, the index is transmitted alongside its quantized amplitude (using a uniform mid-tread quantizer). The cost of encoding an atom index is $\log_2(M)$ bits where $M$ is the size of the dictionary in which the atom was chosen. A quantized approximation $\hat{f}_n$ of $f$ is obtained after $n$ iterations with a characteristic $SNR$ (same as $SRR$ but computed on $\hat{f}_n$ instead of $f_n$) and associated bit-rate.

The signals are taken from the MPEG Sound Quality Assesment Material (SQAM) dataset, their sampling frequency is reduced to 32000 Hz. Three cases are compared in this study:
\begin{description}
 \item[Coarse MP] Matching Pursuit with a union of MDCT basis with no additional parameter. The MDCT basis have sizes from 4 to 512 ms (8 dyadic scales from $2^7$ to $2^{14}$ samples) hop size is $50\%$ in each scale.
 \item[LoMP] Locally Optimized Matching Pursuit (LoMP) with a union of MDCT Basis. This is a locally-adaptive pursuit where an atom is first selected in the coarse dictionary. Then its time localization is optimized in a neighbourhood to find the best local atom in the full dictionary. This pursuit is a slightly suboptimal equivalent of a Matching Pursuit using the full dictionary (Full MP). It is nevertheless less computationally intensive. However, it requires the computation (and the transmission) of an additional parameter (the local time shift) per atom (see \cite{Moussallam2011} for more details).
 \item[RSS MP] MP with a random sequence of subdictionaries. Each subdictionary is itself a union of MDCT basis, each basis of scale $s_k$ is randomly translated in time by a parameter $\tau_k^i \in [-s_k/4:s_k/4]$. It is worth noticing that each scale is independently shifted. The sequence $\mathbf{T} = \{ \tau_k^i \}_{k=1..K, i=1..n}$ is known in advance at both encoder and decoder side, i.e. there is no need to transmit it.
\end{description}
Step SSMP and Jump SSMP are two deterministic SSMP algorithms that are further described in Section \ref{sec:discussion}.

\subsection{Sparsity results}
\begin{figure}
\centering
 \includegraphics[width=9cm]{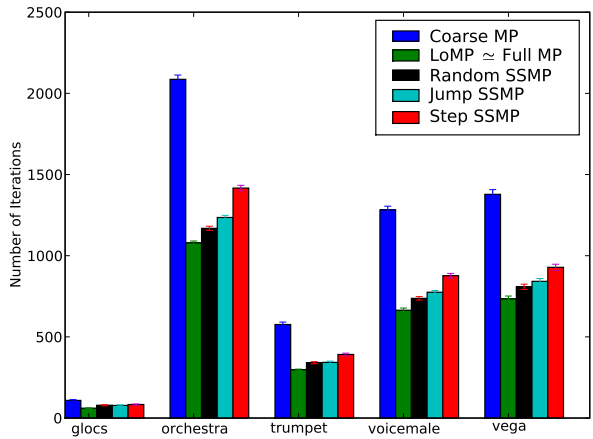}
 \caption{\textit{Number of iterations needed to achieve 10 dB of SRR. Means and standard deviation for various 4 seconds length audio signals from the MPEG SQAM dataset with MP on various fixed and varying sequences of subdictionaries and random initial time offset. Step SSMP and Jump SSMP are two deterministic SSMP.}}
 \label{fig:MultRandom_SRR_10_3xMDCT_100trials_Length_131072_resized}
\end{figure}
Figure \ref{fig:MultRandom_SRR_10_3xMDCT_100trials_Length_131072_resized} shows the number of iterations needed with the three described algorithm (and two variants discussed in \ref{sec:randomvsDeterministic}) to decompose short audio signals from the MPEG SQAM dataset \cite{MPEGDatabase2003} (a glockenspiel, an orchestra, a male voice , a solo trumpet and a singing voice), at 10 dB of SRR. One can verify that RSS MP yields sparser representations (fewer atoms atoms are needed to reach a given SRR) than Coarse MP. This statement remains valid at any given SRR level in this setup.

Experiments where run 100 times with the audio signal being randomly translated in time at each run. Figure \ref{fig:MultRandom_SRR_10_3xMDCT_100trials_Length_131072_resized} shows that empirical variance remains low, which gives us confidence in the fact that RSS MP will give sparser representations in most cases. So far we have not found a natural audio example contradicting this. 

\subsection{Compression results}
Figure \ref{fig:MultRandom_SRR_10_3xMDCT_100trials_Length_131072_resized} shows that the locally-adaptive algorithm (LoMP) is better in terms of sparsity of the achieved representation than RSS MP. However, each atom in these representations is more expensive to encode. Actually, one must transmit an additional local time shift parameter. With RSS MP no such parameter needs to be transmitted.
This gives RSS MP a decisive advantage over the two other algorithms for audio compression.

\begin{figure*}
 \centering
\includegraphics[width=8cm]{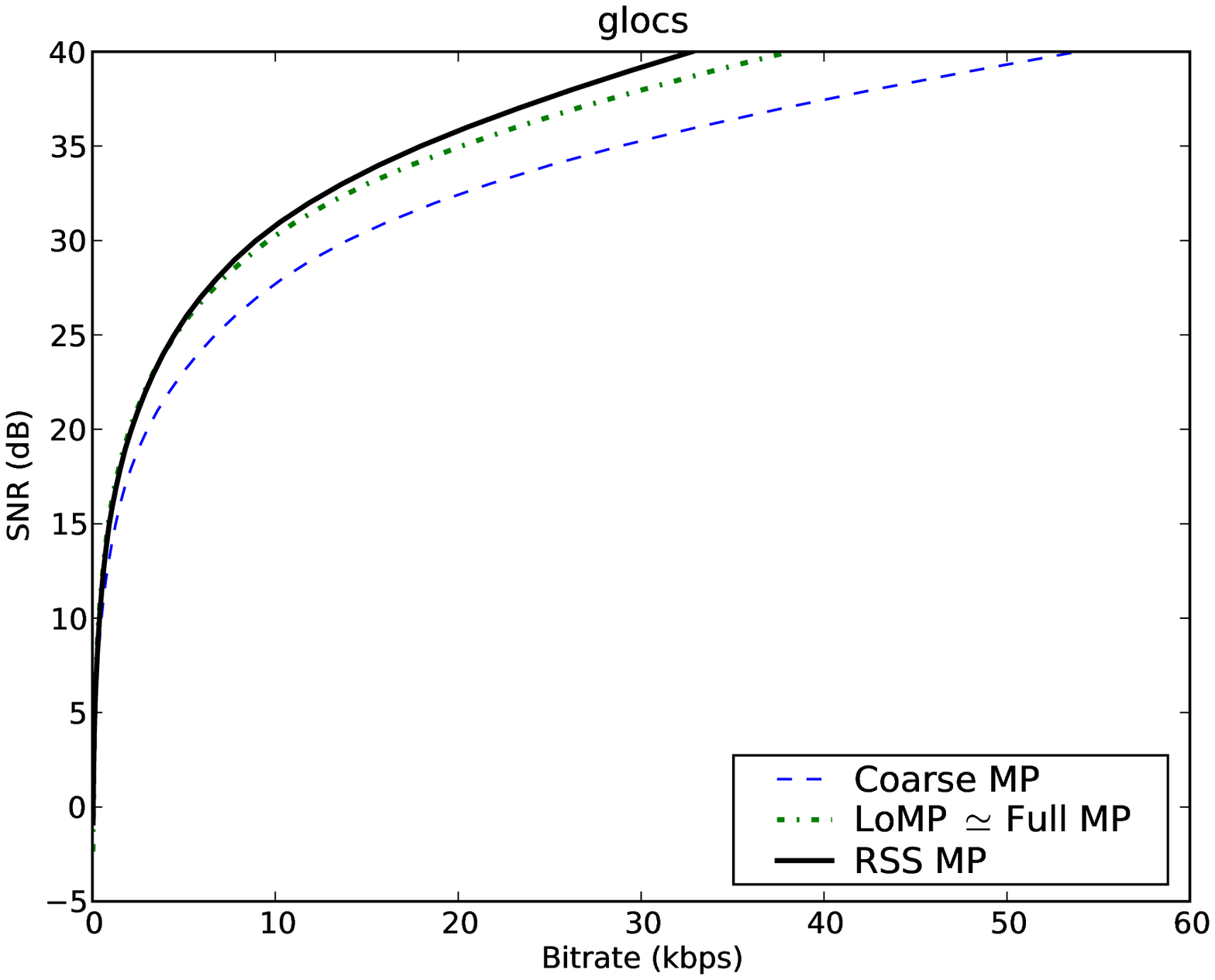}
\includegraphics[width=8cm]{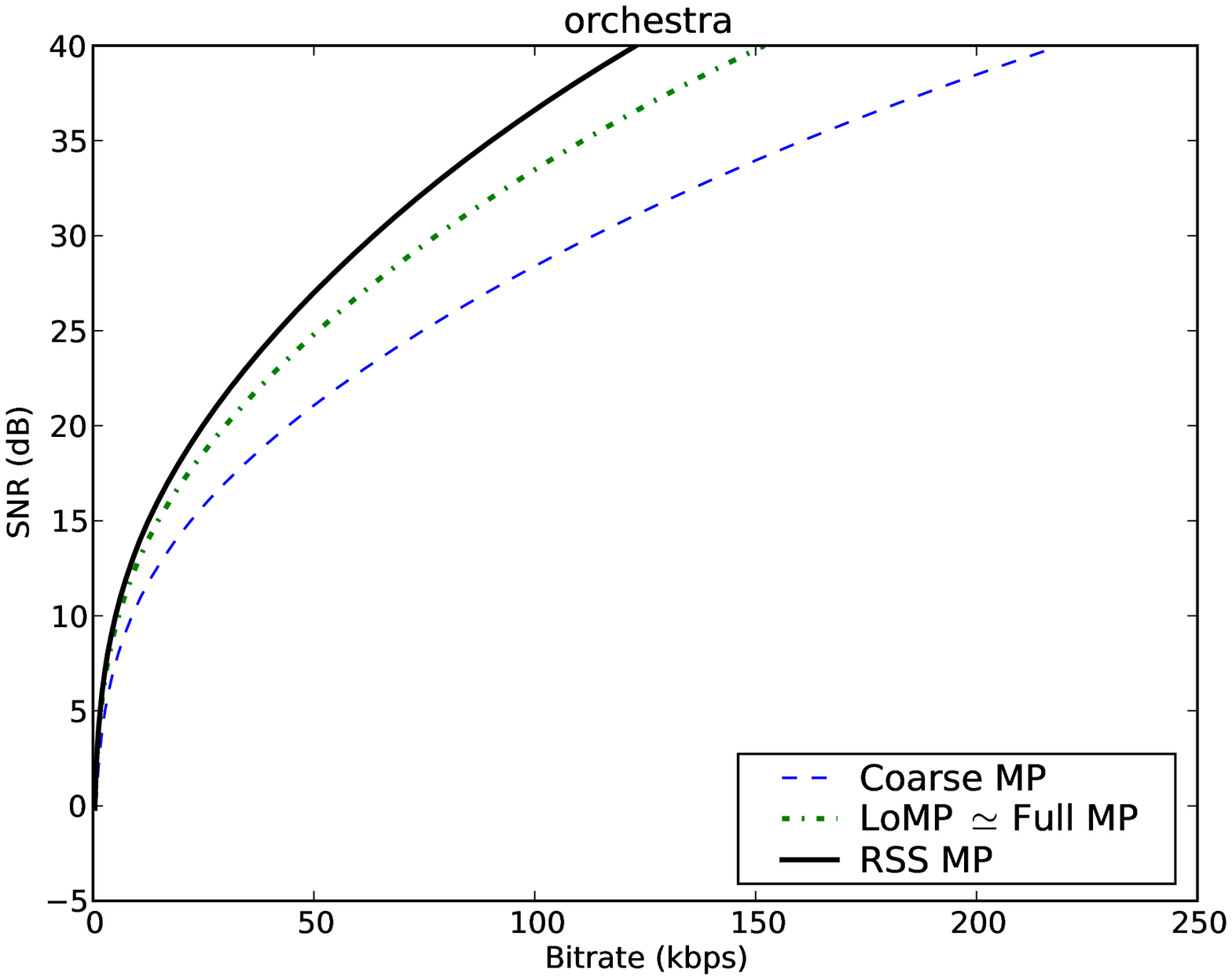}
\includegraphics[width=8cm]{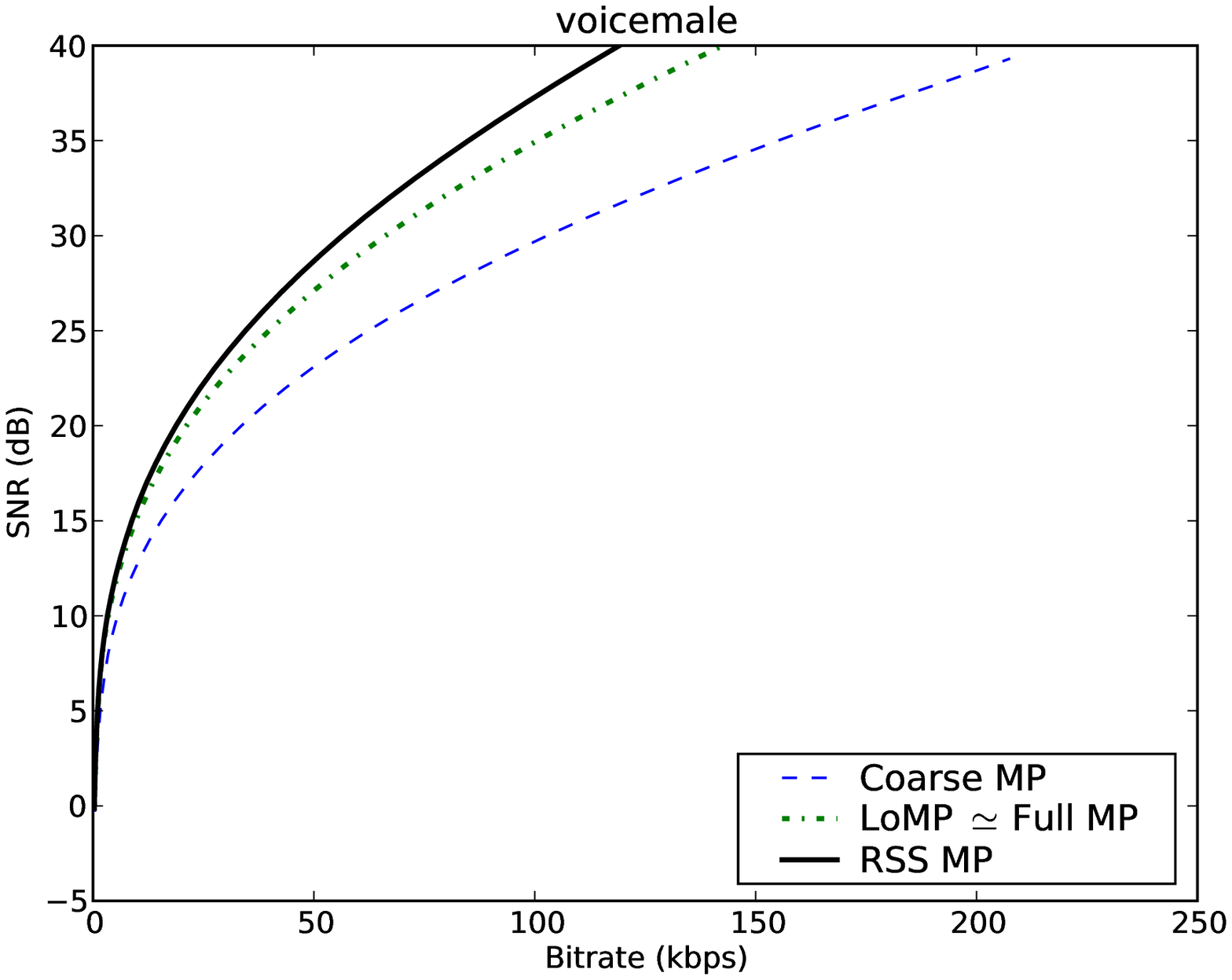}
\includegraphics[width=8cm]{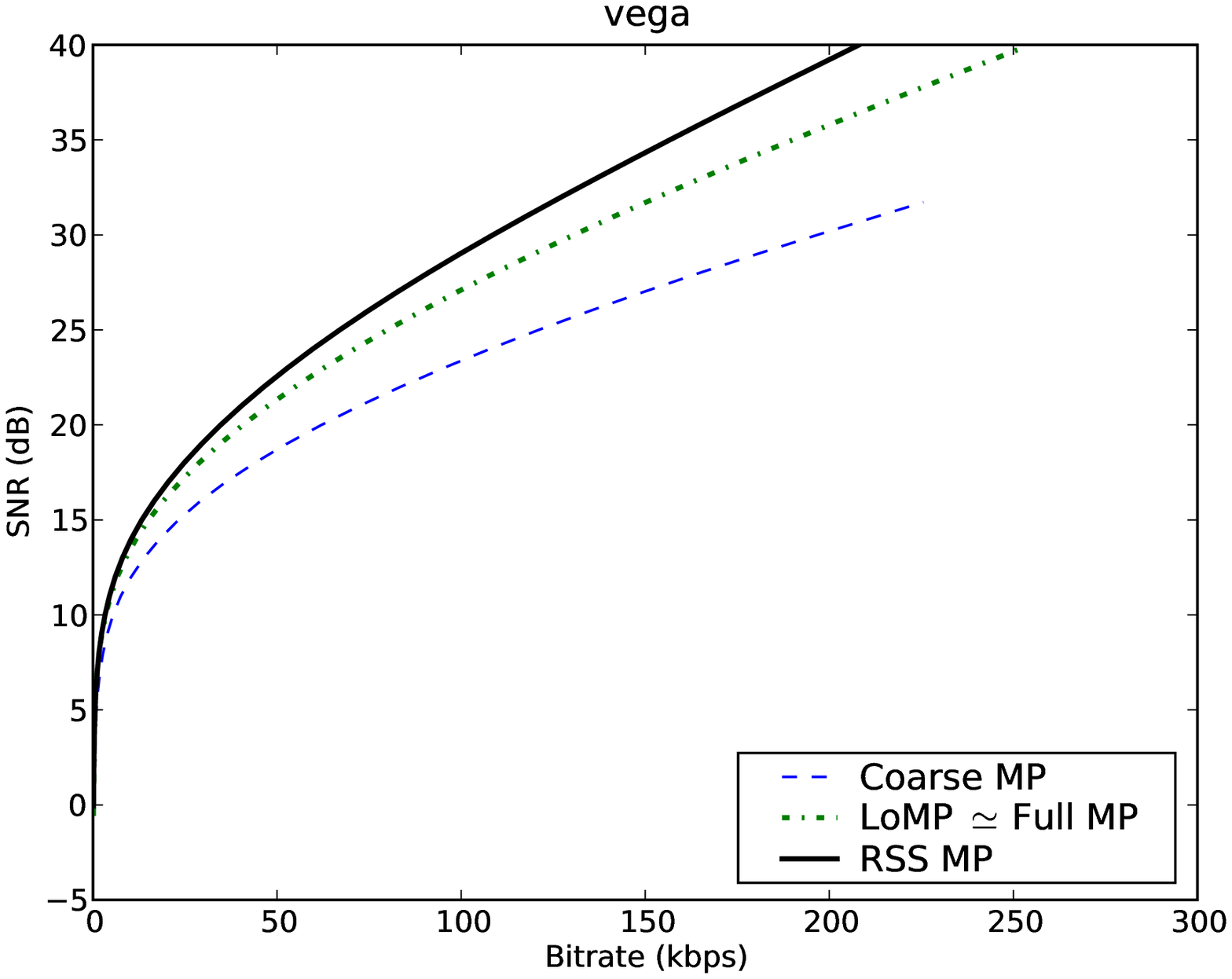}

\caption{\textit{SNR/Bitrate curves for $6s$-length signals from the MPEG SQAM dataset with multi-scale MDCT dictionaries ($2^7$,$2^{10}$ and $2^{13}$ samples per window).}}
 \label{fig:codingCosts}
\end{figure*}

As an example, let us examine the case of the last audio example labeled \textit{vega}. This is the first seconds from Susan Vega's song Tom's Dinner.  The three algorithms are run with a union of 3 MDCT scales (atoms have lengths of 4, 32 or 256 ms). Again these scores have been averaged over 100 runs with the signal being randomly offset at each run and showed very little empirical variance. In order to reach e.g. 20 dB of SRR (i.e before weight quantization):
\begin{itemize}
 \item Coarse MP selected 6886 atoms. If each has a fixed index coding cost of 18 bits and weight coding cost of 16 bits, a total cost of 231 kBits would be needed.
 \item LoMP selected 3691 atoms. With same index and weight cost plus an additional time shift parameter whose cost depends on the length of the selected atom, the total cost would be 149 kBits.
 \item RSS MP selected 3759 atoms (with empirical standard deviation of 16 atoms). Each atom has the same fixed cost as in the Coarse MP case. The average total size is thus 126 kBits.
\end{itemize}
Using an entropy coder as in Figure \ref{fig:Encoding} does not fundamentally change these results. Figure \ref{fig:codingCosts} summarizes compression results with Coarse MP, LoMP and RSS MP for various audio signals from this dataset. Although these comparisons do not use the most efficient quantization and coding tools, it is clear from this picture that the proposed randomization paradigm can be efficiently used in audio compression with greedy algorithms.

In this proof of concept, the quality measure adopted was the $SNR$. However, such algorithms and dictionaries are known to introduce disturbing \textit{ringing} artifacts and pre-echo. We have not experienced that the proposed algorithm increased nor reduced these artifacts compared to the other pursuits. Furthermore, pre-echo control and dark energy management techniques can be applied with this algorithm just as with any other.
Audio examples are available online\footnote{ http://www.tsi.telecom-paristech.fr/aao/?p=531}.

\section{Discussion}
\label{sec:discussion}
Having presented the novel algorithm and an application to audio coding, further questions are investigated in this section. First an experimental study is conducted to illustrate why the sequence of subdictionaries should be pseudo-randomly generated. Then, the RSS scheme is applied to orthogonal pursuit and finally a complexity study is provided.

\subsection{Random vs Deterministic}
\label{sec:randomvsDeterministic}
In the above experiments we have constructed a sequence of randomly varying subdictionaries by designing a sequence of time shifts $\mathbf{T} = \{ \tau_k^i \}_{k=1..K, i=1..n}$ by means of a pseudo-random generator with a uniform distribution: $\forall (k,i) \tau_k^i \sim \mathcal{U}(-s_k/4,s_k/4)$. There are other ways to construct such a sequence, some of which are deterministic. However, we have found that this pseudo-random setup is interestingly the one that gives the best performances.

To illustrate this, let us recall the setup of Section \ref{sec:RMP_Sounds} and compare the sparsity of representations achieved with sequences of subdictionaries built in the following manner:
\begin{description}
 \item[RANDOM:] $\mathbf{T}$ is randomly chosen: $\forall (k,i) \tau_k^i \sim \mathcal{U}(-s_k/4,s_k/4)$
 \item[STEP:] $\mathbf{T}$ is constructed with stepwise increasing sequences : $\forall (k,i) \tau_k^i = mod(\tau_k^{i-1} + 1, s_k/2)$ and $\forall k, \tau_k^0 = 0$. This sequence yields dictionaries that are quite close from one iteration to the next. 
 \item[JUMP:] $\mathbf{T}$ is constructed with \textit{jumps} : $\forall (k,i) \tau_k^i = mod(\tau_k^{i-1} + s_k/4 -1, s_k/2) $and $\forall k, \tau_k^0 = 0$. This sequence yields dictionaries that are very dissimilar from one iteration to the next, but are quite close to dictionaries 2 iterations later.
\end{description}

Figure \ref{fig:MultRandom_SRR_10_3xMDCT_100trials_Length_131072_resized} shows that among all cases, the random strategy is the best one in terms of sparsity. The worst strategy is STEP. This can be explained by the fact that, from one iteration to the next, the subdictionaries are  well correlated. Indeed all the analysis windows are shifted simultaneously by the same (little) offset.
The JUMP strategy is already better. Here the different analysis scales are shifted in different ways. Low correlation between successive subdictionaries appears to be an important factor. Experiments were run 100 times and the hierarchical pattern remain unchanged for higher SRR levels.

Many more deterministic strategies have been tried, the random strategy remains the most interesting one when no further assumption about the signal is made. We explain this phenomenon by noticing that pseudo random sequences are more likely to yield subdictionaries that are very uncorrelated one to another not only on a iteration-by-iteration basis but also for a large number of iterations.

\subsection{Orthogonal pursuits}
\label{sec:OMP}
 \begin{figure}
 \centering
 \includegraphics[width=10cm]{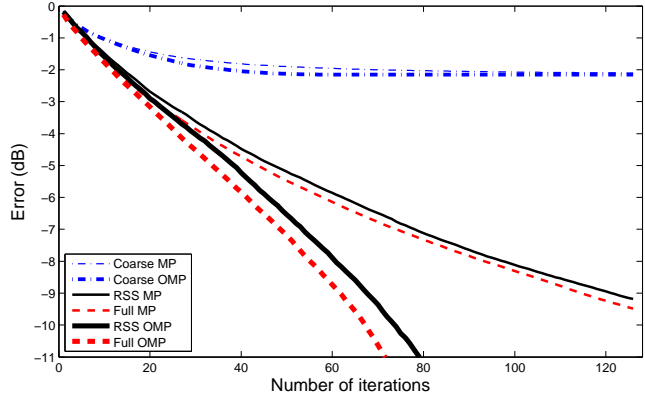}
 \caption{\textit{Comparison between algorithm MP (thin) and OMP (bold) with fixed coarse subdictionary $\Phi_{\mathcal{I}^0}$ (dashed dotted blue), random sequential subdictionaries $\Phi_{\mathcal{I}^n}$ and full dictionary $\Phi$ (dashed red). Input signal is a collection of $N/2$ vectors from $\Phi$ with normal weights. Results averaged over 1000 runs. Here $N = 128$, $M=256$}}
 \label{fig:Convergence_ToyExpeApprox}
\end{figure}
Other SSMP family members can be created with a simple modification of their atom selection rule. In particular, one may want to apply this technique to Orthogonal Matching Pursuit. The resulting algorithm would then be called RSS OMP.

To evaluate RSS OMP, we have recreated the toy experiment of \cite{Blumensath_GP2008} and use 1000 random dictionaries of size $128 \times 256$ with atoms drawn uniformly from the unit sphere. Then, 1000 random signals are created using 64 elements from the dictionaries with unit variance zero mean Gaussian amplitudes. For each signal and corresponding dictionary, the decomposition is performed using MP and OMP with a fixed subdictionary of size $128 \times 64$ (Coarse MP and Coarse OMP), a random sequence of subdictionaries of size $128 \times 64$ (RSS MP and RSS OMP), and the complete dictionary (Full MP and Full OMP). Matlab/Octave code to reproduce this experiment is available online.

Averaged results are shown Figure \ref{fig:Convergence_ToyExpeApprox}. Fixed subdictionary cases (Coarse MP/OMP) show a saturated pattern caused by incompleteness of the subdictionary. Although working on subdictionaries four times smaller, pursuits on random sequential subdictionaries exhibit a behavior close to pursuits on the complete dictionary. Potential gains are even more visible with the OMP algorithm, where the probabilistic parsing of the large dictionary allows for a much better error decay rate than the fixed subdictionary case (Coarse OMP). 


\subsection{Computational Complexities}

\subsubsection{Iteration complexities in the general case}
Let $f$ be in $\mathbb{R}^N$ and $\Phi$ be a redundant dictionary of $M$ atoms also in $\mathbb{R}^N$.  Let $\Phi_{\mathcal{I}}$ be a subdictionary of $\Phi$ of $L$ atoms. Complexities in the general case are given at iteration $i$ by Table \ref{tab:comp_1}. When no update trick is used, Step 1 requires the projection of the $N$ dimensional residual onto the dictionary followed by the selection of the maximum index. Step 2 is rather simple for standard MP, it only involves an update of the residual through a subtraction of the selected atom. For OMP though, a Gram Matrix computation followed by an optimal weights calculation is needed to ensure orthogonality between the residual and the subspace spanned by all previously selected atoms. Complexities of this two additional steps increases with the iteration number and has led to the development of bypassing techniques that limit complexity using iterative QR factorization trick \cite{Blumensath_GP2008}. Complexities in Table \ref{tab:comp_1} are given according to these tricks.

\begin{table}
\centering
\begin{tabular}{l l | c  c  c  c  | c  c }
 & Step & Full MP  & Coarse MP  & Full OMP & Coarse OMP & RSS MP & RSS OMP \\ 
\hline
\textbf{Step 1} & Projection  & $\mathcal{O}(MN)$ & $\mathcal{O}(LN)$ & $\mathcal{O}(MN)$ & $\mathcal{O}(LN)$ & $\mathcal{O}(LN)$  & $\mathcal{O}(LN)$\\
	& Selection & $\mathcal{O}(M)$ & $\mathcal{O}(L)$ & $\mathcal{O}(M)$ & $\mathcal{O}(L)$ & $\mathcal{O}(L)$  & $\mathcal{O}(L)$\\
\hline
\textbf{Step 2} & Gram Matrix  & 0 & 0 & $\mathcal{O}(iN)$ & $\mathcal{O}(iN)$ & $0$ & $\mathcal{O}(iN)$ \\
	& Weights & 0 & $0$ & $\mathcal{O}(i^2)$ & $\mathcal{O}(i^2)$ & $0$ & $\mathcal{O}(i^2)$ \\
	& Residual & $\mathcal{O}(N)$ & $\mathcal{O}(N)$ & $\mathcal{O}(iN)$ & $\mathcal{O}(iN)$ & $\mathcal{O}(N)$ & $\mathcal{O}(iN)$ \\
	\hline
 \multicolumn{2}{c |}{\textbf{Total}}	& $\mathcal{O}(MN)$ & $\mathcal{O}(LN)$ & $\mathcal{O}(i^2 + MN)$ & $\mathcal{O}(i^2 + LN)$ & $\mathcal{O}(LN)$ & $\mathcal{O}(i^2 + LN)$ \\
\end{tabular} 
\caption{Complexities of the different steps for various algorithm in the general case (no update trick for projection). $N$ is the signal dimension. $M$ is the number of atoms in $\Phi$ and $L$ the number of atoms in subdictionaries. The iteration number $i$ is always smaller than $N$.}
\label{tab:comp_1}
\end{table}

\subsubsection{Update tricks and short atoms}
Although RSS MP has the same complexity in the general case as Coarse MP, there are practical limitations to its use. First, Mallat\textit{ et al} \cite{Mallat_TSP1993} proposed a simple update trick that effectively accelerates a decomposition. Projection step is performed using previous projection values and the pre-computed inner products between atoms of the dictionary. Changing the subdictionary on an iteration-by-iteration basis prevents from using this trick. 

An even greater optimization is available when using fast transforms and more specifically when only local updates are possible. One of the main accelerating factor proposed for MP in the MPTK framework \cite{Krstulovic_ICASSP2006} and for OMP in \cite{Mailhe2009_LocOMP} is to limit the number of projections that require an update to match the support of the selected atom. Again, when changing dictionaries at each iteration, this trick can no longer be used to accelerate Step 1. 

Using fast transforms, the complexity of the projection step at first iteration reduces to $\mathcal{O}(M \log N)$ for Full MP \cite{Mailhe2009_LocOMP}. Additionally if one uses short atoms of size $P << N$, it further reduces to $\mathcal{O}(M \log P)$. When local update can be used instead of full correlation, the projection step after iteration 1 has an even more reduced complexity of $\mathcal{O}(\alpha P \log P)$ in $\Phi_{\mathcal{I}}$ where $\alpha = \frac{L}{N}$ is the redundancy factor and of $\beta P \log P$ in $\Phi$ where $\beta = \frac{M}{N} = \alpha \frac{M}{P}$. Regardless of the dictionary, the residual only needs a local update around the chosen atom.

\begin{table}
\centering
\begin{tabular}{l l | c  c | c }
 & Step & Full MP & Coarse MP & RSS MP  \\ 
\hline
\textbf{Step 1} & Projection  & $\mathcal{O}(\beta P \log P)$ & $\mathcal{O}(\alpha P \log P)$ & $\mathcal{O}(\alpha N \log P)$ \\
	& Selection & $\mathcal{O}(\beta P)$ & $\mathcal{O}(\alpha P)$ & $\mathcal{O}(\alpha N)$ \\
\hline
\textbf{Step 2} & Gram Matrix  & 0 & 0 & 0 \\
	& Weights & 0 & $0$ & 0\\
	& Residual & $\mathcal{O}(P)$ & $\mathcal{O}(P)$ & $\mathcal{O}(P)$ \\
\end{tabular} 
\caption{Complexities after iteration 1 of the different steps for various algorithm in structured time-frequency dictionaries with fast transform and limited atomic support $P << N$. In this case, the local update speed up trick can not be applied to RSS MP.}
\label{tab:comp_2}
\end{table}

Table \ref{tab:comp_2} shows that in this context, the proposed modification becomes less competitive with respect to the standard algorithm on fixed subdictionary and can even be slower than the full dictionary case if $ L > \frac{P M}{L}$. However, it can still be considered as a viable alternative since memory requirements for full dictionary projections can get prohibitive.

In order to limit the computational burden, it is possible to change the subdictionary only once every several iterations. If the subdictionary $\Phi_{\mathcal{I}^i}$ remains unchanged for $J$ consecutive iterations, then the usual tricks for speeding up the projection can be applied. When $\Phi_{\mathcal{I}^i}$ changes at iteration $i+J$, all projections need be recomputed. Although it can accelerate the algorithm, one may expect a resulting loss of quality.


\subsubsection{Sparsity/Complexity tradeoff}
Another way to tackle the complexity issue is by choosing an appropriate size for the subdictionaries. Since RSS MP yields much sparser solutions with dictionaries of the same size than Coarse MP, we can further reduce the size of the subdictionaries in order to accelerate the computation while still hoping to get compact representations. Indeed, in the discussion above, we have only compared complexities per iteration, but the total number of iterations is also important.

 \begin{figure}
 \centering
 \includegraphics[width=10cm]{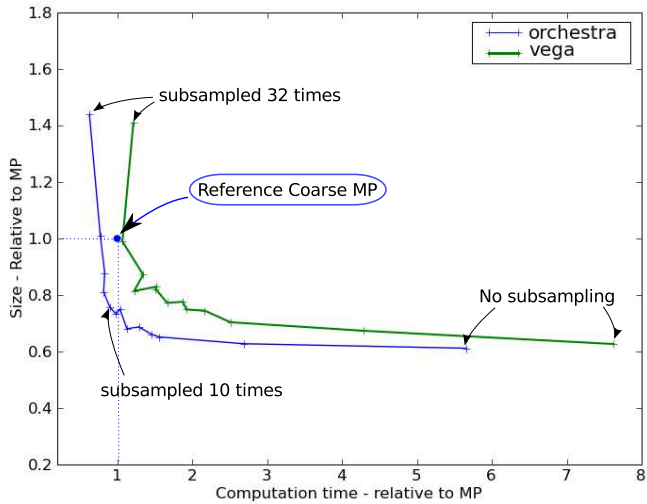}
 \caption{\textit{Illustration of the sparsity/complexity tradeoff that can be achieved by controlling the size of the subdictionaries in RSS MP.  }}
 \label{fig:Compromise2_2_Signals_BitrateSpeed}
\end{figure}

To illustrate this idea, we have used subsampled subdictionaries with a subsampling factor varying from 1 (no subsampling) to 32. The subsampling is performed by limiting the frame indexes in which we calculate the MDCT projections. It is important to notice that the subdictionaries are no longer redundant, and not even complete when the subsampling factor is greater than 2. Figure \ref{fig:Compromise2_2_Signals_BitrateSpeed} shows for two signals how the subsampling impacts the sparsity of the representation (here expressed as a bit-rate calculated as in section \ref{sec:RMP_Sounds}) and the computational complexity relative to the reference Coarse MP algorithm for which we have implemented the local update trick. 

Dictionaries used in this setup are multiscale MDCT with 8 different scales (from 4 to 512 ms), the experiment is run 100 times on a dual core computer up to a SRR of 10 dB. Using random subdictionaries 10 times smaller than the fixed one of Coarse MP, we can still have good compression (  cost 30 \% less than the reference) with the computation being slightly faster.

More interestingly, this subsampling factor, associated with the RSS MP algorithm, controls a sparsity/complexity tradeoff that allows multiple use cases depending on needs and resources.

\section{Conclusion}
In this work, we proposed a modification of greedy algorithms of the MP family. Using a pseudo random sequence of subdictionaries instead of a fixed subdictionary can yield sparser approximations of signals. This sparsity basically comes at no additional coding cost if the
random sequence is known in advance, thus giving our algorithm a clear advantage for compression purposes as shown here for audio signals. On the downside, the modification may increase complexity, but we have proposed a tradeoff strategy that helps reducing computation times.

The idea presented in this work can be linked to existing techniques from other domains, where randomness is used to enhance signal processing tasks. Dithering for quantization is one such technique. Another example is spread spectrum in communication \cite{Dixon1994}, where a signal is multiplied by a random sequence before being transmitted on a bandwidth that is much larger than the one of the original content. Here, encryption is more important a goal than compression. However, performing RSS MP of a signal somehow requires the definition of a key (the pseudo-random sequence) that must be known at the reception side in order to decode the representation. Moreover, the representation may also live in a much larger space than the original content. It is worth noticing for instance the work of Puy \textit{et al} \cite{Puy2011} where the authors apply spread spectrum techniques to the compressed sensing problem. 

Other experiments should be conducted for feature extraction tasks, since the good convergence properties indicates that the selected components carry meaningful information. For instance, music indexing \cite{Ravelli2009} and face pattern recognition \cite{Phillips1998} use Matching Pursuits to derive the features that serve for their processing. Further work will have to investigate whether RSS MP can improve on the performance of other algorithms on such tasks.   

\section*{Acknowledgments}
 We are very grateful to the anonymous reviewers for their detailed comments that helped improve the quality of this paper. We also would like to thank Dr. Bob Sturm for the interest he showed in our work and his relevant remarks and advice. Discussing with him was both an opportunity and a pleasure. This work was partly supported by the QUAERO programme, funded by OSEO, French State Agency for innovation.

\bibliographystyle{plain}
\bibliography{RMPBiblio}

\end{document}